\documentclass[preprint,showpacs,preprintnumbers,amsmath,amssymb]{revtex4}

\usepackage{graphicx}
\usepackage{dcolumn}
\usepackage{bm}

\begin{document}
\vspace{1cm}
\begin{center}
{\large \bf{ Amplified biochemical oscillations in cellular systems}} 

\vspace{0.5cm}

{\textbf{A.~J. McKane$^{1}$, J.~D. Nagy$^{2,3}$, 
T.~J. Newman$^{3,4}$, and M.~O. Stefanini$^{4}$}} \\
\bigskip
{\small\emph{
$^1$Theoretical Physics, School of Physics and Astronomy,
University of Manchester, Manchester M13 9PL, UK \\
$^2$Life Science Department, Scottsdale Community College, Scottsdale AZ
85256\\
$^3$School of Life Sciences, Arizona State University, Tempe AZ 85287\\
$^4$Department of Physics and Astronomy, Arizona State 
University, Tempe AZ 85287}}
\end{center}

\vspace{2cm}

\begin{center}
{\bf {Abstract}}
\end{center}
We describe a mechanism for pronounced biochemical oscillations,
relevant to microscopic systems, such as the intracellular
environment. This mechanism operates for reaction schemes which, when
modeled using deterministic rate equations, fail to exhibit
oscillations for any values of rate constants. The mechanism relies on
amplification of the underlying stochasticity of reaction kinetics
within a narrow window of frequencies. This amplification allows
fluctuations to ``beat the central limit theorem,'' having a dominant
effect even though the number of molecules in the system is relatively
large. The mechanism is quantitatively studied within simple models
of self-regulatory gene expression, and glycolytic oscillations.

\newpage

\section{Introduction}
\label{intro}
Biochemical oscillations have been studied from complementary
experimental and theoretical perspectives for many years. They appear
to be generic processes in biological systems, with numerous examples
known in both epigenetic and metabolic contexts
[1-4].  Well known instances are
circadian rhythms (\emph{e.g.} in microorganisms
[5,6]), and the oscillation of ATP and ADP
concentrations during phosphorylation of fructose-6-phosphate (F6P), a
key step in glycolysis [7,8]. The study of biochemical
oscillations has benefited from an intimate collaboration between
experimentalists and theorists: reaction networks are pieced together
in the laboratory, with hypotheses severely constrained by results
from mathematical models [4,8]. These mathematical
models are generally constructed using deterministic chemical rate
equations. Within this modeling framework, the signature of
oscillatory behavior is the existence of a limit cycle. The absence of
a limit cycle is assumed to imply that the underlying reaction scheme,
upon which the model is based, does not have cyclic
behavior. Recently, a number of groups have challenged this assumption
[9-12], mainly in the context of calcium
oscillations.  They have studied deterministic rate equations which
support a limit cycle over a certain range of parameter values, and
have found, from computer simulations, that the addition of noise can
expand the region of parameter space in which cycles occur. These
authors demonstrate that this effect can be maximized for particular
values of the system size or noise strength. On this basis, the effect
has been termed ``internal noise stochastic resonance.'' We show here
that internal noise can have a far more profound effect on biochemical
reaction kinetics; namely, inducing amplified oscillations in systems
which, when modeled with rate equations, lack a limit cycle throughout
their entire parameter space. Thus, from the view of conventional
deterministic modeling, these reaction schemes would be immediately
ruled out as candidates to describe biochemical oscillations. 

In this paper we develop a theoretical framework, using the Van Kampen
system-size expansion [13-15], which allows an 
exact analytic description of these stochastically induced 
cycles -- indeed, our theory reduces to a problem in linear algebra 
regardless of the reaction scheme under study, thereby allowing 
straightforward analysis. A crucial point is that this effect operates only 
in systems composed of a relatively modest number of molecules -- typically 
in the range $10^{2}-10^{6}$. Many biochemical reactions within cells
operate with numbers of molecules in this range, which leads us to
believe that intracellular processes may well exploit this
amplification mechanism.  Our result is at odds with the intuitive
notion that fluctuations can be safely ignored for systems composed of
many thousands of molecules -- the reason being that, here, the
amplitude of fluctuations is composed of two factors: the usual
statistical factor $1/\sqrt{N}$ (where $N$ is the typical number of
molecules in the system), and a large factor $R \gg 1$ arising from
the amplification of the underlying noise. If $R/\sqrt{N} \sim O(1)$
or larger, the intuitive conclusion that fluctuations can be ignored
for $N \gg 1$, is incorrect.

For ease of presentation we shall introduce this mechanism through two
relatively simple examples -- one epigenetic (gene regulation), 
the other metabolic (glycolysis).
Generalizations to more complex reaction schemes is straightforward,
since, as already mentioned, the analysis of fluctuations reduces to
an exercise in linear algebra. We will develop the theoretical ideas
using the language of the gene regulation model (defined in Figure 1a)
for concreteness, because it involves only the populations of two
constituents, unlike the model of glycolytic oscillations which
involves four.

The outline of the remainder of this paper is as follows. In section
II we introduce the urn model representation of the gene regulation
model, and proceed to instantiate this as a master equation. The
deterministic limit of the master equation is derived in section III,
and is shown to correspond to the usual chemical rate equation
description.  We calculate the effect of weak fluctuations in section
IV, using the Van Kampen system-size expansion. We obtain a
description of the fluctuations as a set of \emph{linear} stochastic
differential equations, which can therefore be analyzed exactly. The
power-spectra obtained from these equations have resonance peaks which
considerably enhance the $1/\sqrt{N}$ effects we would naively expect
from fluctuations. The analysis (master equation, deterministic
limit, first-order fluctuations) is repeated in a condensed form
for Selkov's model of glycolysis in section V. Again, we find a
strong peak in power spectrum of the fluctuations indicating 
amplified oscillations. We end with a discussion of our results in
the broader context of biochemical reaction networks.

\section{Individual-based stochastic model for the gene regulation model}
\label{IBM}
Consider, first, a simple reaction scheme in which an enzyme inhibits
the transcription of its parent gene (Figures 1a and 1b). More
elaborate reaction schemes based on this simple model have long been
proposed to explain circadian rhythms [5,16]. It is
well-known that a deterministic model of this simplest reaction scheme
involving two chemical agents (mRNA and the enzyme) will fail to
produce cycles -- at least three chemical agents (the mRNA, the
enzyme, and an intermediate form, e.g. the primary RNA transcript) are
required. It is important to recall that the deterministic model is an
approximate representation of the actual chemical kinetics -- it is
strictly accurate for an infinite bath of molecules which are
well-mixed. In order to describe a finite number of molecules it is
necessary to use a discrete stochastic formulation in which one tracks
the probability distribution of chemical concentrations over time. We
have developed a standard stochastic treatment of the two-agent
reaction scheme in Figures 1a and 1b.  The calculational steps used in
the approach are illustrated in a flow-chart (Figure 2). By taking the
limit of the number of molecules in our system, $N$, to be infinite,
we indeed recover the deterministic theory -- this is an important
benchmark. To handle the case of a finite number of molecules, we
perform a system size expansion [13], which is a standard
technique from the theory of stochastic processes, in which
fluctuations are accounted for within a perturbative treatment. The
second order terms in this expansion describe the fluctuations about
the deterministic theory. These fluctuations satisfy linear equations
and their statistics can be solved exactly. In particular, we
calculate the power spectrum of these fluctuations. We find that for a
wide range of parameters, the power spectrum has a pronounced maximum
within a narrow window of frequencies.

In order to perform systematic numerical comparisons between
integrating the rate equations and simulating the stochastic model it
is convenient to envisage the reactions (Figures 1a and 1b)
occurring within two baths, as shown in figure 3a. The empty state
$\emptyset$ has been replaced by null constituents $E_1$ and $E_2$ to
be discussed further below. The reason for introducing two baths is
simply because it is easier to set up the dynamics of the system and
is closer in spirit to other systems (grounded in population dynamics)
which have been analyzed in a similar way, and are well-understood
[14,15].

Our aim is to build an individual-based stochastic model, and so the
basic ingredients will be the number of $M$ (mRNA) and $E_1$ (null)
constituents in bath 1 and of $P$ (enzyme) and $E_2$ (null)
constituents in bath 2 --- in a given stochastic realization at a
given time $t$. The dynamics will consist of picking constituents from
the baths at each time step and attempting the specified
reactions. Performing many runs of such reactions will enable us to
collect a large number of realizations and extract average
behavior. The reactions proceed according to the rates shown in Figure
3a, and if the selection of constituents does not correspond to one of
the four reactions, then the molecules are returned to their
respective baths without any action being taken.

On the basis of the specification so far, and summarized in Figure 3a,
numerical simulations of the model can be carried out. We use the
elegant Gillespie algorithm [17] which uses the
information encoded in the reaction scheme to generate random time
increments, in each of which a randomly selected reaction is forced to
occur. The time increments and reactions are selected according to
weighted distributions in such a way that the probability distribution
of the stochastic time series generated is exact. The results of some
of these simulations are displayed in Figures 4 and 5.

It is also possible to derive a set of equations which describe the
stochastic process which is defined by the model we have specified. To
do this, we first note that there has been an implication that the
choice of constituents at a given time step is a random process, only
dependent on the numbers of the various molecules in the baths at that
time, and not on choices or availability at previous time
steps. Assumptions of this kind imply that the process is Markov, and
so can be modeled using a master equation
[13,18]. This is essentially a continuous time
version of a Markov chain. Before we can write down this equation, we
need to define some more quantities.

Let us denote the number of molecules of the various kinds as follows:
in bath 1 the number of molecules of $M$ is $n_1$ and in bath 2 the
number of molecules of $P$ is $n_2$. The state of the system is then
denoted by the two numbers $(n_{1},n_{2})$. We will frequently write
this as $\underline{n}$ when we simply want to refer to the general
state of the system. Note that we do not confer the status of
independent variables on the numbers of $E_1$ or $E_2$
constituents. If we denote the total number of constituents in bath 1
by $N_1$ and that in bath 2 by $N_2$, then the number of $E_1$ and
$E_2$ constituents are simply what is required to make up these
numbers: $(N_{1}-n_{1})$ $E_1$ and $(N_{2}-n_{2})$ $E_2$ constituents.
This is why it was necessary to introduce the null constituents: if
they had not been included, the number of $M$ and $P$ molecules would
not have the freedom to vary --- they provide room for independent
changes in the numbers of both $M$ and $P$ molecules.

We may now define transition rates from one state, $\underline{n}$, to
a different state $\underline{n}'$. For instance, in the fourth
reaction, the number of $P$ molecules increases by 1 (recall that the
number of $E_2$ molecules is not an independent variable). So in this
case, $\underline{n} = (n_{1},n_{2})$ and $\underline{n}' =
(n_{1},n_{2}+1)$. We denote the transition rate by $T
(\underline{n}'|\underline{n})$. In our convention, initial states are
on the right and final states on the left. When picking the
constituents, there is a probability of $n_1/N_1$ of $M$ being chosen,
$(N_{2}-n_{2})/N_{2}$ that an $E_2$ is chosen, and the reaction
happens at a rate $k_2$. This gives the result $k_{2} (n_1/N_1)
(N_{2}-n_{2})/N_{2}$ for this particular transition rate. Others may
be found in the same way. A complete listing with $\underline{n} =
(n_{1},n_{2})$ is
\begin{itemize}
\item[1.] \ $M \stackrel {\mu_1}{\longrightarrow } E_1$, 
\ $\underline{n}'= (n_{1}-1,n_{2})$,
\begin{displaymath}
T (\underline{n}'|\underline{n})= \mu_{1} \frac{n_1}{N_1}\,. 
\end{displaymath}
\item[2.] \ $P \stackrel {\mu_2}{\longrightarrow } E_2$, 
\ $\underline{n}'= (n_{1},n_{2}-1)$,
\begin{displaymath}
T (\underline{n}'|\underline{n})= \mu_{2} \frac{n_2}{N_2}\,. 
\end{displaymath}
\item[3.]\ $E_{1} \stackrel {k_1}{\longrightarrow} M$,  
\ $\underline{n}'= (n_{1}+1,n_{2})$,
\begin{displaymath}
T (\underline{n}'|\underline{n})= k_{1} \frac{(N_{1}-n_{1})}{N_1}
= k_{0}\exp{\left( - \lambda n_{2}/N_{2} \right)}\,\frac{(N_{1}-n_{1})}{N_1}\,.
\end{displaymath}
\item[4.]\ $M + E_{2} \stackrel {k_2}{\longrightarrow} M + P$,
\ $\underline{n}'= (n_{1},n_{2}+1)$,
\begin{displaymath}
T (\underline{n}'|\underline{n})= k_{2} \frac{n_1}{N_1} 
\frac{(N_{2}-n_{2})}{N_2}\,.
\end{displaymath}
\end{itemize}
Note, we use an exponential function $\exp(-\lambda {\rm [P]})$ to
model the down-regulation of transcription. The primary reason is this
form introduces one extra parameter ($\lambda $) into the model,
unlike the typical Hill form [16] $(1+\alpha {\rm
[P]}^{q})^{-1}$ which introduces two new parameters ($\alpha$ and
$q$). Furthermore, in deterministic studies (albeit using more complex
three-component models), it is found that large values of $q$ ($\sim
10$) are required for limit cycles to occur [16]. For such
large values of $q$, the Hill form is a rapidly decaying function, and
it seems reasonable to replace it by a simple exponential function.

Having defined the transition rates, we are now in a position to write
down the master equation. It has the general form
[13,18]
\begin{equation}
\frac{d}{dt} P(\underline{n}, t) = 
\sum_{\underline{n}' \neq \underline{n}}\,
T(\underline{n}|\underline{n}') P(\underline{n}', t) -
\sum_{\underline{n}' \neq \underline{n}}\,
T(\underline{n}'|\underline{n}) P(\underline{n}, t)\,,
\label{gen_master}
\end{equation}
where $P(\underline{n}, t)$ is the probability that the system is in
the state $\underline{n}$ at time $t$. This equation has a simple
interpretation: the first term on the right hand side is the sum of
the transition rates into the state $\underline{n}$ from all other
states $\underline{n}'$ and the second term is the sum of the
transition rates out of the state $\underline{n}$ into all other
states $\underline{n}'$. When the second term is subtracted from the
first, it gives the rate of change of the probability $P$.

So far we have formulated a discrete stochastic model and given
specified analytic forms for transition probabilities between states,
which can be used in conjunction with the master equation
(\ref{gen_master}) to, in principle, solve for the probability,
$P(\underline{n}, t)$, that the system is in state $\underline{n}$ at
time $t$. We will now start from the master equation and (i) determine
the form that the model takes in the mean-field limit, and (ii)
implement the system-size expansion to study the origin of the cycling
behavior, which is absent in the deterministic (mean-field) limit.

\section{The deterministic limit}
\label{deter}
A straightforward way of obtaining the deterministic version of the
model, valid in the limit of very large system size, from the master
equation is to multiply (\ref{gen_master}) by $n_{1}$ and $n_2$ in
turn and to sum over all the states. This generates rate equations
which are the deterministic equations if correlations between the
variables are ignored. Let us illustrate the method in the case of
$n_2$. We wish to calculate $\langle n_{2}
\rangle=\sum_{\underline{n}} n_{2}P(\underline{n}, t)$ by multiplying
the master equation by $n_2$ and summing over $\underline{n}$. On the
left-hand side we find $d\langle n_{2} \rangle/dt$. On the right-hand
side are two terms which are nearly equal and opposite --- if it was
not for the $n_2$ factor, they would be. The only difference between
the two terms is that $\underline{n}$ and $\underline{n}'$ are
interchanged. So if $n_2$ does not change in a reaction, as for
reactions 1 and 3, then the two contributions do in fact cancel
out. If it decreases by 1, as in reaction 2, then a shift in the sum
over $n_2$ gives an overall contribution of $-1$, and if it increases
by 1, as in the reaction 4, then a shift in the sum over $n_2$ the
other way, gives an overall contribution of $+1$. This gives the
result
\begin{displaymath}
\frac{d}{dt} \langle n_{2} \rangle = 
- \mu_{2} \left\langle \frac{n_2}{N_2} \right\rangle 
+ k_{2} \left\langle \frac{n_1}{N_1} \frac{(N_{2}-n_{2})}{N_2} \right\rangle\,.
\end{displaymath}
So far this is exact. The mean-field approximation enters through
ignoring correlations, which vanish as $N_{1},N_{2} \to \infty$. So,
for example, this means that $\langle n_{i} n_{j} \rangle =\langle
n_{i} \rangle \langle n_{j} \rangle$ for $i, j=1,2$. If we make this
approximation and introduce the fractions of $M$ and $P$ to be
$\phi_{1}$ and $\phi_{2}$ respectively, in the limit $N_{1},N_{2} \to
\infty$, then we find
\begin{displaymath}
\frac{d\phi_{2}}{dt} = - \frac{\mu_2}{N_2} \phi_{2} + \frac{k_2}{N_2}\phi_{1}
\left( 1 - \phi_{2} \right)\,.
\end{displaymath}
This is the required deterministic equation. So in summary, if we
define $\phi_{i} = n_{i}/N_{i}$ and scale the time by introducing
$\tau = t/N_1$, then the deterministic equations corresponding to the
individual based stochastic model we have defined are
\begin{eqnarray}
\frac{d\phi_{1}}{d\tau} & = & - \mu_{1} \phi_{1} + k_{0} 
\exp{\left( - \lambda \phi_{2} \right)} \left( 1-\phi_{1} \right)\,, 
\label{one} \\
\sigma^{-1}\,\frac{d\phi_{2}}{d\tau} & = & - \mu_{2} \phi_{2} + k_{2} \phi_{1} 
\left( 1-\phi_{2} \right)\,,
\label{two} 
\end{eqnarray}
where $\sigma = N_{1}/N_{2}$. Note that the mean-field approximation
as applied to the first equation implies that $\langle \exp{\left( -
\lambda n_{2}/N_{2} \right)} \rangle = \exp{\left( - \lambda \langle
n_{2} \rangle /N_{2} \right)}$.

Most of the theoretical investigations of biochemical reactions start
from a set of differential equations of the type (\ref{one}) and
(\ref{two}).  One of the first quantities of interest in such studies
are the fixed points of the system. If we denote the fixed points with
an asterisk and define $X = \lambda \phi^{*}_{2}$, then $X$ satisfies
the transcendental equation
\begin{equation}
k_{0}\,e^{-X} = \frac{\mu_{1} \mu_{2} X}{k_{2}\lambda - (k_{2}+\mu_{2}) X}\,,
\label{eqn_for_X}
\end{equation}
with the fixed points for $\phi_{1}$ given by
\begin{equation}
\phi^{*}_{1} = \frac{\mu_{2} X}{k_{2} (\lambda - X)}\,.
\label{phi_1_fp}
\end{equation}
Note that $1 - \phi^{*}_{1} > 0$ and so the denominator of the
right-hand side of Eq.~(\ref{eqn_for_X}) is never zero. Also, since
the left-hand side of this equation is monotonically decreasing from
its value at $X=0$ and the right-hand side is monotonically increasing
from its value at $X=0$, it follows that there is always a unique
solution for $X$ and therefore always just one fixed point. The
condition $0 \leq \phi^{*}_{2} \leq 1$ implies that $0 \leq X \leq
\lambda$, but the condition $0 \leq \phi^{*}_{1} \leq 1$ gives the
stronger constraint
\begin{equation}
X \leq \frac{k_{2} \lambda}{k_{2}+\mu_{2}}\,.
\label{X_constraint}
\end{equation}

The stability of these fixed points would always be of interest, but
the question takes on an added significance in our case, since the
stability matrix associated with the non-trivial fixed point
(Eqs.~(\ref{eqn_for_X}) and (\ref{phi_1_fp})) plays a central role in
the analysis of the cycling phenomenon. To determine it, let us write
the mean-field equations (\ref{one}) and (\ref{two}) in the form
$d\phi_{i}/d\tau=f_{i} (\underline{\phi})$ where $i=1,2$. The fixed
points are found from solving $f_{i} (\underline{\phi})=0$ and a
linear stability analysis consists of writing
$\phi_{i}=\phi_{i}^{*}+(\hat{\phi}_{i}/\sqrt{N_i})$, where the
$\hat{\phi}_{i}$ are small deviations from the fixed point
[19] .  Note that we have included extra factors of
$1/\sqrt{N_i}$ in the small deviations from the fixed point --- which
simply amounts to a re-scaling of $\hat{\phi}_{i}$ --- compared to the
usual linear stability analysis, so as to make contact with the
system-size expansion in the next section. Linearizing about the fixed
point gives $d\hat{\phi}_{i}/d\tau = \sum_{j} M_{ij} \hat{\phi}_{j}$,
where $M_{ij}$ is defined by $M_{ij} = \partial f_{i}/\partial
\phi_{j}|_{\rm FP}$. Here FP means ``evaluated at the fixed
point''. The explicit forms for the entries of the matrix $M$ are
\begin{eqnarray}
M_{11} &=& - \mu_{1} - k_{0}\exp{ \left( - \lambda \phi_{2}^{*} \right)}\,, 
\nonumber \\
M_{12} &=& - k_{0} \lambda \sigma^{1/2} 
\exp{ \left( - \lambda \phi_{2}^{*} \right)} \left( 1-\phi_{1}^{*} \right)\,,
\nonumber \\
M_{21} &=& k_{2} \sigma^{1/2} \left( 1-\phi_{2}^{*} \right)\,,
\nonumber \\
M_{22} &=& - \sigma \mu_{2} - \sigma k_{2} \phi_{1}^{*}\,. 
\label{entries_of_M}
\end{eqnarray}
These entries may be rewritten using the fixed-point equations to give
\begin{eqnarray}
M_{11} &=& - \frac{\mu_1}{1-\phi^{*}_{1}}\,, \ \ \ 
M_{12} = - \lambda \mu_{1} \sigma^{1/2}\,\phi^{*}_{1}\,, \nonumber \\
M_{21} &=& \mu_{2} \sigma^{1/2}\,\frac{\phi^{*}_{2}}{\phi^{*}_{1}}\,, \ \
M_{22} = - \sigma k_{2} \frac{\phi^{*}_{1}}{\phi^{*}_{2}}\,.
\label{fpentries_of_M}
\end{eqnarray}
From these expressions it is clear that all entries of $M$ have a
definite sign: $M_{11}, M_{12}$ and $M_{22}$ are negative, and
$M_{21}$ is positive. Therefore, the determinant and the trace of $M$
are positive and negative respectively for all parameter choices. This
already tells us that the fixed point is stable, but further analysis
is required if we wish to know whether or not the the fixed point is
approached in an oscillatory fashion. This will be discussed again in
the next section.

\section{Analysis of the fluctuations}
\label{analysis}
In contrast with the innate discreteness of the stochastic model, the
mean-field equations involve functions of continuous variables. This
is one of the reasons why they are more amenable to analytic
treatment. It is the limit $N_{1},N_{2} \to \infty$ which leads to the
continuity of the mean-field equations, as well as to the elimination
of the fluctuations. In the section we will discuss how we can keep
continuity, but still not lose the stochastic nature of the
system. This is achieved by using new continuous variables $x_{1},
x_{2}$ in place of the previously used discrete variables $n_{1},
n_{2}$ to describe the probability distribution. The explicit form of
the replacements are
\begin{equation}
\frac{n_1}{N_1} = \phi_{1} + \frac{x_1}{\sqrt{N_1}}\,, \ \ \  
\frac{n_2}{N_2} = \phi_{2} + \frac{x_2}{\sqrt{N_2}}\,.
\label{replacements}
\end{equation}
The $1/\sqrt{N}$ terms are present since, by the central-limit
theorem, we expect fluctuations to be of the order of $1/\sqrt{N}$
when the variables $n_{1},n_{2}$ are expressed in terms of the
fractions $n_{1}/N_1$ and $n_{2}/N_2$. As $N_{1}, N_{2} \to \infty$,
these fluctuations vanish, and the system is entirely described by the
mean-field variables $\phi_{i} (\tau)$, which can be found, in
principle, by solving Eqs.~(\ref{one}) and (\ref{two}) with given
initial conditions. If we imagine a plot of the probability
distribution $P$ along the vertical axis, and the variables $n_i$
along the horizontal axes, for various values of $\tau$, then at the
initial time $\tau=0$ $P$ is a delta-function spike at the starting
values of the $n_i$. As $\tau$ increases, not only does the position
of the peak move, but the probability distribution also broadens due
to fluctuations. The $\phi_{i} (\tau)$, which are the solutions of
Eqs.~(\ref{one}) and (\ref{two}), tell us the position of the peak of
the distribution and the variables $x_{i} (\tau)$ tell us something
about the distribution itself. Actually it turns out that to order
$1/\sqrt{N}$ (here we use $N$ to mean $N_1$ or $N_2$, since they are
assumed to be of the same order), the probability distribution is
Gaussian, and since the position of the peak is specified, all that is
left to determine is the variance. Higher terms in the $1/\sqrt{N}$
expansion give deviations from the Gaussian form, which our numerical
simulations show are very small for reasonable values of $N$. Once the
leading (deterministic) contributions have been subtracted out from
(\ref{replacements}) (by, in effect, continuously moving the origin to
the peak of the probability distribution) only fluctuations remain. If
the $1/\sqrt{N_1}$ and $1/\sqrt{N_2}$ terms are then factored out, we
may once again take $N \to \infty$. In this way the $x_i$ become
continuous variables, and terms of different orders can be identified
in the master equation.

The actual implementation of the system-size expansion is
straightforward, if tedious. It is discussed clearly in
[13] and in some detail in [14]. We shall
illustrate it by applying it to one term in the master equation
only. The reader should be able to understand the essential features
of the method from this, and can then consult the above references to
get a broader picture. Let us take reaction 2, in which the number of
$P$ molecules decreases by 1. It gives a contribution to the term
$T(\underline{n}'|\underline{n}) P(\underline{n}, t)$ in the master
equation (\ref{gen_master}) which is equal to $\mu_{2} (n_2/N_2)
P(n_{1}, n_{2}, t)$. It also gives a contribution to
$T(\underline{n}|\underline{n}') P(\underline{n}' , t)$ in the same
equation which is equal to $\mu_{2} ([n_{2}+1]/N_{2})P(n_{1}, n_{2}+1,
t)$. We can combine these two terms as follows:
\begin{equation}
{\rm Reaction\ 2\ }: \ \ \ \left( {\cal E}_{2} - 1 \right) 
\left[ \mu_{2} \frac{n_2}{N_2} P(n_{1}, n_{2}, t) \right]\,,
\label{reaction2}
\end{equation}
where ${\cal E}_{2}$ is a step operator which is defined in terms of
its action on functions of the $n_i$ by ${\cal E}_{2}^{\pm 1} \psi
(n_{1}, n_{2}) = \psi (n_{1}, n_{2}\pm 1)$. Similar operators can be
defined for the other variables. The advantage of using these
operators is that, within the replacement scheme (\ref{replacements}),
they have a simple form for large $N$. For example,
\begin{equation}
{\cal E}_{2}^{\pm 1} = 1 \pm \frac{1}{\sqrt{N_2}}\frac{\partial}
{\partial x_{2}} + \frac{1}{2N_2} \frac{\partial^2}
{\partial x_{2}^{2}} + \ldots\,.
\label{step_op}
\end{equation}
Substituting (\ref{replacements}) and (\ref{step_op}) in
(\ref{reaction2}), and expanding in inverse powers of $\sqrt{N_2}$ one
finds a contribution proportional to $1/\sqrt{N_2}$:
\begin{equation}
\frac{1}{\sqrt{N_2}}\,\mu_{2} \phi_{2}\,\frac{\partial \Pi}{\partial x_2}\,,
\label{leading}
\end{equation}
and a term of order $1/N_{2}$:
\begin{equation}
\frac{1}{N_2}\,\mu_{2} \left[ \frac{\partial}{\partial x_2} 
\left( x_{2} \Pi \right) + \frac{1}{2}\,\phi_{2} \frac{\partial^{2} \Pi}
{\partial x_{2}^{2}} \right]\,.
\label{next_to_leading}
\end{equation}
There are higher order terms, but this is as far as we need to go in
the expansion. The quantity $\Pi$ which appears in
Eqs.~(\ref{leading}) and (\ref{next_to_leading}) is numerically equal
to $P$, but is instead a function of the $x_i$ and of $t$. This means
that the left-hand side of (\ref{gen_master}) now reads
\begin{eqnarray}
\frac{\partial P}{\partial t} &=& 
\frac{\partial \Pi}{\partial t} - \sqrt{N_1} \frac{d\phi_1}{dt} 
\frac{\partial \Pi}{\partial x_1} - \sqrt{N_2} \frac{d\phi_2}{dt} 
\frac{\partial \Pi}{\partial x_2} \nonumber \\
&=& \frac{1}{N_1} \frac{\partial \Pi}{\partial \tau} - 
\frac{1}{\sqrt{N_1}} \frac{d\phi_1}{d\tau} 
\frac{\partial \Pi}{\partial x_1} - \sigma^{-1} \frac{1}{\sqrt{N_2}} 
\frac{d\phi_2}{d\tau} \frac{\partial \Pi}{\partial x_2}\,. 
\label{lhs_master}
\end{eqnarray}
Equating the right-hand side of the master equation
(Eqs.~(\ref{leading}) and (\ref{next_to_leading})) to the left-hand
side (Eq.~(\ref{lhs_master})) order by order, we find that
$\sigma^{-1} d\phi_{2}/d\tau$ has a contribution $- \mu_{2} \phi_{2}$
(the $\partial \Pi/\partial x_2$ cancel out) and
\begin{equation}
\frac{\partial \Pi}{\partial \tau} = \mu_{2} \sigma 
\left[ \frac{\partial}{\partial x_2} 
\left( x_{2} \Pi \right) + \frac{1}{2}\,\phi_{2} \frac{\partial^{2} \Pi}
{\partial x_{1}^{2}} \right] + \ldots\,,
\label{start_of_FPE}
\end{equation}
where the dots mean that the reactions other than 2 will also give a
contribution.

This partial, but explicit, calculation allows us to see more clearly
how the expansion works. At leading order we have a contribution to
the mean-field equation for $\sigma^{-1} d\phi_{2}/d\tau$ which is
equal to $-\mu_{2} \phi_2$, which indeed appears in Eq.~(\ref{two}),
and in none of the other mean-field equations. Therefore, an
alternative, and in some sense more systematic, way of obtaining the
mean-field equations is as the leading order terms in the large-$N$
expansion. The next-to-leading terms give a partial differential
equation for the probability distribution $\Pi (\underline{x}, t)$
which, when we include all the other reactions, has the form
\begin{equation}
\frac{\partial \Pi}{\partial \tau} = - \sum_{i}\,\frac{\partial}{\partial x_i} 
\left( A_{i} (\underline{x})\,\Pi \right) + \frac{1}{2} \sum_{i,j}\,
B_{ij} \frac{\partial^{2} \Pi}{\partial x_{i} \partial x_{j}}\,.
\label{FP_eqn}
\end{equation}
From (\ref{start_of_FPE}) we see that 
$A_{2} (\underline{x}) = - \mu_{2} \sigma x_{2} +\ldots$ 
and $B_{11} = \mu_{2} \sigma \phi_{2} + \ldots$. In fact when all the 
reactions are included, the $A_{i} (\underline{x})$ remain linear functions 
of the $x_j$ and the $B_{ij}$ remain independent of them. We may therefore 
write
\begin{equation}
A_{i} (\underline{x}) = \sum^{2}_{j=1} M_{ij} x_{j}\,.
\label{AandM}
\end{equation}
This means that the probability distribution at next-to-leading order,
$\Pi(\underline{x}, \tau)$, is completely determined by two $2 \times
2$ matrices: $M$ and $B$, whose elements are independent of the $x_j$,
and only functions of the $\phi_j$. In fact it is a characteristic of
the large-$N$ expansion that $M$ is nothing else but the Jacobian
matrix with elements $\partial f_{i}/\partial \phi_j$, which may be
calculated directly from the mean-field equations (\ref{one}) and
(\ref{two}). If the initial transients have died away and the solution
to the deterministic equations has approached a fixed-point
$\underline{\phi}^{*}$, then $M$ will simply be the stability matrix
at that fixed point. The stability matrix for the fixed point
(\ref{eqn_for_X}) and (\ref{phi_1_fp}) is given in the previous
section. The matrix $B$ has entries
\begin{eqnarray}
B_{11} &=& \mu_{1} \phi_{1} + k_{0}\exp{ \left( - \lambda \phi_{2} \right)} 
\left( 1-\phi_{1} \right)
\nonumber \\
B_{22} &=& \mu_{2} \sigma \phi_{2} + k_{2} \sigma \phi_{1} 
\left( 1-\phi_{2} \right) \nonumber \\
B_{12} &=& B_{21} = 0\,.
\label{entries_of_B}
\end{eqnarray}
  
So, to summarize the position so far, we know the $M_{ij}$ and
$B_{ij}$ in terms of the parameters which define the
individually-based stochastic model, and therefore by solving
Eq.~(\ref{FP_eqn}) we can find all we need to know about the
fluctuations for large $N_1$ and $N_2$. The partial differential
equation (\ref{FP_eqn}) is a Fokker-Planck equation --- a continuous
version of the master equation --- and can be solved, in principle,
given the initial condition that $\Pi$ is a delta-function spike at
$\tau=0$. In fact, for the simple case where the $B_{ij}$ are
independent of $x_j$ and the $A_i$ are linear functions of the $x_j$,
it can be solved exactly [13].  The result is a
multi-variate Gaussian with $\langle x_{i} \rangle = 0$, as already
mentioned.

Our main aim in this paper is to understand oscillations and for this
one of the main tools is Fourier analysis. The form of
Eq.~(\ref{FP_eqn}) is not so useful for this purpose, but fortunately
there is a completely equivalent formulation of the stochastic process
which is ideally suited to investigation using Fourier
transforms. Rather than write an equation for the probability
distribution function $\Pi$, an equation for the actual stochastic
variables $x_{i} (\tau)$ can be given, in other words, the problem may be
formulated as a set of stochastic differential equations of the Langevin type.
The Langevin equations which are equivalent to (\ref{FP_eqn}) are
[13]
\begin{equation}
\frac{d x_i}{d\tau} = A_{i} (\underline{x}) + \eta_{i} (\tau)\,,
\label{Langevin}
\end{equation}
where $\eta_{i} (\tau)$ is a Gaussian noise with zero mean and with a 
correlation function given by
\begin{equation}
\langle \eta_{i} (\tau) \eta_{j} (\tau ') \rangle = B_{ij} 
\delta (\tau - \tau ' )\,.
\label{noise_corr}
\end{equation}
The system defined by Eqs.~(\ref{Langevin}) and (\ref{noise_corr}) is
ideally suited to Fourier analysis, since the equations
(\ref{Langevin}) are linear (since the $A_i$ are) and
Eq.~(\ref{noise_corr}) implies that the noise is white, that is, the
Fourier transform of its correlation function is
frequency-independent.

Taking the Fourier transform of (\ref{Langevin}) gives
\begin{equation}
-i \omega \tilde{x}_{i} (\omega) = \sum^{2}_{j=1} M_{ij} 
\tilde{x}_{j} (\omega) + \tilde{\eta}_{i} (\omega)\,,
\label{Langevin_FT}
\end{equation}
where the tilde denotes the Fourier transform. We may write this as 
\begin{equation}
\sum^{2}_{j=1} \Phi_{ij} (\omega) \tilde{x}_{j} (\omega) = 
\tilde{\eta}_{i} (\omega)\,, \ \ \ \Phi_{ij} (\omega) \equiv 
-i \omega \delta_{ij} - M_{ij}\,.
\label{Phi}
\end{equation}
The Fourier transform of $\eta_{i} (t)$ has the correlation function
\begin{equation}
\langle \tilde{\eta}_{i} (\omega) \tilde{\eta}_{j} (\omega ' ) \rangle = 
B_{ij} (2\pi) \delta( \omega + \omega ' )\,.
\label{correlation}
\end{equation}
From Eq.~(\ref{Phi}) we obtain 
$\tilde{x}_{i}(\omega)=\sum_{j}\Phi^{-1}_{ij}(\omega)\tilde{\eta}_{j}(\omega)$,
and averaging the squared modulus of $\tilde{x}_{i}$ gives the power-spectra
\begin{equation}
P_{i} (\omega) = \langle |\tilde{x}_{i} (\omega)|^{2} \rangle =
\sum^{2}_{j=1}\,\sum^{2}_{k=1}\,\Phi^{-1}_{ij} (\omega) B_{jk} 
\left( \Phi^{\dag} \right)^{-1}_{ki} (\omega)\,,
\label{power_defn}
\end{equation}
where we have used $\Phi_{ij} (-\omega) = \Phi^{\dag}_{ji} (\omega)$.
[We have omitted the proportionality factor $2\pi \delta (0)$. In
practice, when comparing the analytically calculated power spectra to
those generated from a numerical time series, one uses a discrete
Fourier transform, and the proportionality factor is simply equal to
the time increment of the recorded values in the time series.]

To isolate the resonance, we note that $P_{i} (\omega)$ has the form
of the ratio of two power series in $\omega$. The denominator, which
will largely control the position of the resonance, can be simply
expressed as the determinant of the matrix $\Phi$. If we define
$D(\omega)= \det \Phi(\omega)$, then the denominator is just
$|D(\omega )|^2$.

In Eqs.~(\ref{Langevin_FT})-(\ref{power_defn}) we have written
everything in a rather general form, since the formalism can be seen
to generalize to the case of an arbitrary number of constituents
rather easily, such as in the model of glycolysis in the next section
which has 4 constituents. However, for the model we are presently
considering, there are only two constituents and the expressions for
$P_{i} (\omega),\ i=1,2$ have simple explicit forms:
\begin{equation}
P_{i} (\omega) = \langle |\tilde{x}_{i} (\omega)|^{2} \rangle = 
\frac{\alpha_{i} + \beta_{i} \omega^{2}}{|D(\omega)|^{2}}\,,
\label{power_defn_2}
\end{equation}
where $D(\omega)=-\omega^{2} + i\omega\,{\rm tr} M + \det M$ and where
\begin{eqnarray}
\alpha_{1} &=& B_{11} M^{2}_{22} + B_{22} M^{2}_{12}\,, \ \ \ 
\beta_{1}=B_{11}\,, \nonumber \\
\alpha_{2} &=& B_{11} M^{2}_{21} + B_{22} M^{2}_{11}\,, \ \ \ 
\beta_{2}=B_{22}\,.
\label{alphas_betas}
\end{eqnarray}
Since the elements of the $M$ and $B$ matrices are known in terms of
the parameters of the model and the fixed point values of the
$\phi_{i}$, so are the $\alpha_{i}$ and $\beta_{i}$.

The denominator of the power spectrum in Eq.~(\ref{power_defn_2}) is
given by
\begin{eqnarray}
|D(\omega)|^{2} &=& \left( \omega^{2} - \det M \right)^{2} + 
\left( {\rm tr} M \right)^{2} \omega^{2} \nonumber \\
&=& \left( \omega^{2} - \Omega^{2}_{0} \right)^{2} + \Gamma^{2} \omega^{2}\,,
\label{denom_2}
\end{eqnarray}
where, since in Section \ref{deter} we found $\det M > 0$ and ${\rm
tr} M < 0$, we have introduced $\Omega^{2}_{0} = \det M$ and $\Gamma =
- {\rm tr} M$. So, we may write the power spectrum as
\begin{equation}
P_{i} (\omega) = \langle |\tilde{x}_{i} (\omega)|^{2} \rangle = 
\frac{\alpha_{i} + \beta_{i} \omega^{2}}{\left[ \left( \omega^{2} - 
\Omega^{2}_{0} \right)^{2} + \Gamma^{2} \omega^{2} \right]}\,.
\label{power_defn_p}
\end{equation}
This form for the power-spectrum shows clearly the existence of a
resonance: for a value of $\omega^{2}$ the denominator becomes small,
and the power spectrum has a large peak centered on this frequency.

To analyze the nature of the resonance in more detail, let us set
$z=\omega^2$ and ask for what values of $z$ the power spectrum
(\ref{power_defn_p}) has a maximum. The condition $dP/dz=0$ gives
\begin{equation}
\beta z^{2} + 2\alpha z + \left[ \alpha \left( \Gamma^{2} - 
2 \Omega^{2}_{0} \right) - \beta \Omega^{4}_{0} \right] = 0\,,
\label{extremum_eqn}
\end{equation}
where we have dropped the index $i$ on $\alpha_i$ and $\beta_i$. Let
us begin by neglecting the term $\beta_{i} \omega^2$ in the numerator
of Eq.~(\ref{power_defn_p}) which may be justified for some parameter
choices.  Then the condition (\ref{extremum_eqn}) simply becomes $z =
(2\Omega^{2}_{0} - \Gamma^{2})/2$. Since we require $z=\omega^{2}>0$,
this implies $2\Omega^{2}_{0} > \Gamma^{2}$. In terms of the stability
matrix $M$, this condition reads $2 \det M > ({\rm tr} M)^{2}$, which
implies that the eigenvalues of $M$ are complex. In the last section
we showed that the eigenvalues of the stability matrix were either
real and negative or complex with a negative real part. The condition
that the power spectrum has an extremum imposes the latter.

In reality, the presence of the factor $\beta_{i} \omega^{2}$ in the
numerator of the power spectrum may have a significant effect, and the
full condition (\ref{extremum_eqn}) has to be used. From
Eqs.~(\ref{entries_of_B}) and (\ref{alphas_betas}) we see that the
$\alpha_{i}$ and $\beta_{i}$ are always positive, and so from
Eq.~(\ref{extremum_eqn}) we see that the sum of the roots of this
equation is negative and so at least one of the roots is negative. For
the other one to be positive we require
\begin{equation}
\alpha \left( \Gamma^{2}-2\Omega^{2}_{0} \right) - \beta \Omega^{4}_{0} < 0\,.
\label{full_cond}
\end{equation}
This is the general condition for an extremum of the power spectrum to
occur.  It goes beyond the the previous condition, which only involved
the eigenvalues of the matrix $M$, since it contains the $\beta_i$
which come from consideration of the fluctuations about the fixed
point. We may now ask if the extremum is a maximum. It is
straightforward to show by calculating $d^{2}P/dz^2$ that, if a
positive solution to Eq.~(\ref{extremum_eqn}) exists, then it is
automatically a maximum.

A peak in the power spectrum indicates a resonant
frequency, and a corresponding oscillation at this frequency. [We must
stress that our use of the word ``resonance'' here follows the more
general usage; namely, a peak in the power spectrum as a function of
frequency.  In ``stochastic resonance'' and its derivatives
[9,11] (and references therein), the ``resonance'' refers to a
maximized response of the system as a function of some parameter, such
as the strength of the internal noise. Because of the potential for
confusion, we often use the term ``amplification'' for the effect
under consideration in this work.]  The height of the peak indicates
the strength of the amplification.  The width of the peak indicates
the amount of frequency dispersion one would observe about the
resonant frequency. We have calculated the power spectra exactly for
this epigenetic model. For a specific set of parameters, we show the
power spectra for fluctuations in mRNA and enzyme concentrations
(Figures 4c and 4d respectively). We have also computed stochastic
realizations of time series for this reaction, using the exact
Gillespie algorithm [17] (Figures 4a, b). Note, the
resonant amplification $R$ for mRNA (defined as the height of the peak
of the power spectrum relative to $P(0)$) is quite pronounced ($R \sim
30$), and, indeed, regular oscillatory peaks are easily discernible in
the corresponding time series (Figure 4a). The constant dashed lines
in Figures 4a, b are the results from the deterministic theory --
which predicts a complete absence of cycling. 
We emphasize that there are two competing
quantities which combine to determine the importance of these
amplified oscillations -- these are i) the relative height of the peak
in the power spectrum $R$, and ii) the mean number of molecules in the
system $N$. The amplitude of the oscillations in a time series
measurement of the concentration will be of the order
$R/\sqrt{N}$. This is why the effect dies away for macroscopic systems
in which $N$ is extremely large. In the example given here $R \sim 30$
(for mRNA fluctuations), and so we can estimate that the oscillations
will be negligible in intracellular systems for which $N_{\rm mRNA}
\gg 1000$. Given the simplicity of this two component model, and the
modest estimate above, this mechanism may be relevant to circadian
rhythms in procaryotic cells, such as cyanobacteria [5],
since such cells are small, and have no nuclear membrane separating
transcription and translation processes (justifying, to some degree, a
two-component model).

In Figure 5 we compare the exact form for the mRNA power spectrum,
given above in (\ref{power_defn_2}), to a numerically generated power
spectrum.  The latter is computed from discrete Fourier transforms of
a large number (10,000) of long (60,000 iterations) time series
generated using the exact Gillespie algorithm described earlier. There
is excellent agreement between the exact and numerical spectra, as
expected.

\section{Selkov's model of glycolysis}
\label{selkov}
We now turn to an example from intracellular metabolic dynamics.
Consider the key step in glycolysis in which the enzyme PFK1 catalyzes
the phosphorylation of F6P. This reaction is actually rather complex,
and rather sophisticated models of this process have been developed
over the years, guided by ever more quantitatively precise experiments
[3,7,8].  For the purposes of illustration, we
shall concentrate on one of the first models of this reaction due to
Sel'kov [20]. Although this model is no longer regarded as
being an accurate representation of this reaction, it captures the
essence of the process. Sel'kov's reaction scheme is illustrated in
Figures 1c and 1d. A key parameter in this scheme is the number of ADP
molecules (conventionally denoted by $\gamma $) required to activate
the PFK1 enzyme.  Within the deterministic modeling framework it has
been shown that $\gamma > 1$ is required for cyclic behavior to emerge
from the model (even though the biochemistry of PFK1 demands
$\gamma=1$), and that even then, the cycling exists in a very narrow
range of parameter values. We have reformulated Sel'kov's reaction
scheme using the stochastic framework, but have restricted our
attention to the more biologically plausible case of $\gamma =1$.

As before, the deterministic model is retrieved when the number of
molecules in the system is taken to be infinitely large, and predicts
constant, non-cycling, concentrations.  When this number is finite, we
account for the fluctuations in the system using the system size
expansion, and solve the simple linear theory in order to calculate
the power spectra for the various chemical agents.

The reactions which comprise the Selkov model of a key stage
(phosphorylation of F6P by the PFK1 enzyme) in glycolysis are shown
schematically in Figures 1c and 1d.  Writing them out in a slightly 
different format, they read
\begin{displaymath}
ADP + PFK1 \buildrel {k_3} \over
{\underset{k_{-3}}{\rightleftharpoons}} PFK1/ADP
\end{displaymath}
\begin{displaymath}
\buildrel {\nu_{1}} \over {\longrightarrow} ATP
\end{displaymath}
\begin{displaymath}
ATP + PFK1/ADP \buildrel {k_1} \over
{\underset{k_{-1}}{\rightleftharpoons}} PFK1/ADP/ATP \buildrel
{k_2} \over {\longrightarrow} PFK1/ADP + ADP
\end{displaymath}
\begin{displaymath}
ADP \buildrel {\nu_{2}} \over {\longrightarrow}
\end{displaymath}
Once again we will introduce two baths.

For notational simplicity we will denote ATP by $S_1$, ADP by $S_2$,
PFK1 by $E_2$, PFK1/ADP by $A$ and PFK1/ADP/ATP by $B$. The first two
will be placed in bath number 1, and the last three in bath number 2
(Figure 3b).  It will also be necessary to introduce ``nulls'' into
bath 1, which we will denote by $E_1$. These represent the aqueous
space which can potentially be filled by ATP or ADP. With this
notation $A$ is defined to be $E_{2} S_{2}$ and $B$ is defined to be
$S_{1} E_{2} S_{2}$ and the five reactions are
\begin{itemize}
\item[1.]$E_{1} \buildrel {\nu_{1}} \over {\longrightarrow} S_{1}$
\item[2.]$S_{2} \buildrel {\nu_{2}} \over {\longrightarrow} E_{1}$
\item[3.]$S_{1} + A \buildrel {k_1} \over
{\underset{k_{-1}}{\rightleftharpoons}} B + E_{1}$
\item[4.]$B + E_{1} \buildrel{k_2} \over {\longrightarrow} A + S_{2}$
\item[5.]$S_{2} + E_{2} \buildrel {k_3}
\over{\underset{k_{-3}}{\rightleftharpoons}} A + E_{1}$\,.
\end{itemize}
Let us denote the number of molecules of the various kinds as follows:
in bath 1, the number of molecules of $S_{\alpha},\,\alpha=1,2$ is
$m_{\alpha}$, and in bath 2, the number of molecules of $A$ is $n_1$
and of $B$ is $n_2$.  The state of the system is then denoted by the
four numbers $(m_{1},m_{2},n_{1},n_{2})$, which we will again write as
$\underline{n}$ when we simply want to refer to the general state of
the system. If, as before, we denote the total number of constituents
in bath 1 by $N_1$ and that in bath 2 by $N_2$, then the number of
$E_1$ and $E_2$ constituents is simply $N_{1}-m_{1}-m_{2}$ and
$N_{2}=n_{1}-n_{2}$, respectively.

The transition rates for this model are:
\begin{itemize}

\item[1.]  $\underline{n}'= (m_{1}+1,m_{2},n_{1},n_{2})$.
\begin{displaymath}
T (\underline{n}'|\underline{n})= \nu_{1}
\frac{(N_{1}-m_{1}-m_{2})}{N_1}\,.
\end{displaymath}

\item[2.]  $\underline{n}'= (m_{1},m_{2}-1,n_{1},n_{2})$.
\begin{displaymath}
T (\underline{n}'|\underline{n})= \nu_{2} \frac{m_2}{N_1}\,.
\end{displaymath}

\item[3.] Forward reaction: $\underline{n}'=
(m_{1}-1,m_{2},n_{1}-1,n_{2}+1)$.
\begin{displaymath}
T (\underline{n}'|\underline{n})= k_{1} \frac{m_1}{N_1}
\frac{n_1}{N_2}\,.
\end{displaymath}

\smallskip

Backward reaction: 
$\underline{n}'= (m_{1}+1,m_{2},n_{1}+1,n_{2}-1)$.
\begin{displaymath}
T (\underline{n}'|\underline{n})= k_{-1} 
\frac{(N_{1}-m_{1}-m_{2})}{N_1} \frac{n_2}{N_2}\,.
\end{displaymath}

\item[4.]  $\underline{n}'= (m_{1},m_{2}+1,n_{1}+1,n_{2}-1)$.
\begin{displaymath}
T (\underline{n}'|\underline{n})= k_{2} 
\frac{(N_{1}-m_{1}-m_{2})}{N_1} \frac{n_2}{N_2}\,.
\end{displaymath}

\item[5.] Forward reaction: 
$\underline{n}'= (m_{1},m_{2}-1,n_{1}+1,n_{2})$.
\begin{displaymath}
T (\underline{n}'|\underline{n})= k_{3} \frac{m_2}{N_1} 
\frac{(N_{2}-n_{1}-n_{2})}{N_2}\,.
\end{displaymath}

\smallskip

Backward reaction: 
$\underline{n}'= (m_{1},m_{2}+1,n_{1}-1,n_{2})$.
\begin{displaymath}
T (\underline{n}'|\underline{n})= k_{-3} 
\frac{(N_{1}-m_{1}-m_{2})}{N_1} \frac{n_1}{N_2}\,.
\end{displaymath}

\end{itemize}

To find the deterministic equation we define $s_{\alpha} =
m_{\alpha}/N_1$, $a=n_{1}/N_2$ and $b=n_{2}/N_2$ and scale the time by
introducing $\tau=t/N_1$. This gives the deterministic equations,
corresponding to the individual based stochastic model we have
defined, to be
\begin{eqnarray}
\frac{ds_{1}}{d\tau} & = & \nu_{1}(1-s_{1}-s_{2}) 
-k_{1}s_{1}a + k_{-1}(1-s_{1}-s_{2})b\,, \label{first}\\
\frac{ds_{2}}{d\tau} & = & -\nu_{2}s_{2}+k_{2}(1-s_{1}-s_{2})b
-k_{3}s_{2}(1-a-b) + k_{-3}(1-s_{1}-s_{2})a\,, \label{second}
\\
\sigma^{-1}\,\frac{da}{d\tau} & = & -k_{1}s_{1}a + 
k_{-1}(1-s_{1}-s_{2})b + k_{2}(1-s_{1}-s_{2})b \nonumber 
\\
& & +k_{3}s_{2}(1-a-b)-k_{-3}(1-s_{1}-s_{2})a\,, 
\label{third} \\
\sigma^{-1} \frac{db}{d\tau} & = & k_{1}s_{1}a - k_{-1}
(1-s_{1}-s_{2})b-k_{2}(1-s_{1}-s_{2})b\,, \label{fourth}
\end{eqnarray}
where $\sigma = N_{1}/N_{2}$.  The equations
(\ref{first})-(\ref{fourth}) are Selkov's equations
[8,20], except for the additional factors of
$(1-s_{1}-s_{2})$ which arise from the introduction of the nulls
$E_1$.

To investigate the fixed point structure, let us first note that if we
replace the term $(1-s_{1} - s_{2})$ by $\Omega$ in the equations
(\ref{first})-(\ref{fourth}), it will enable us to examine the Selkov
model ($\Omega=1$) and our deterministic model $\Omega = 1-s_{1} -
s_{2}$, in tandem.  From Eqs.~(\ref{first}) and (\ref{fourth}) we see
that the fixed point has either to have $\Omega^{*}=0$ or $\nu_{1} =
k_{2} b^{*}$ (all fixed point values are again denoted by
asterisks). The first condition cannot hold in the original Selkov
model, but can in our version of the model: in this case we see from
Eq.~(\ref{second}) that $s_{2}^{*}=0$, and so $s_{1}^{*}=1$. This
implies $a^{*}=0$ and $b^{*}$ is indeterminate. If $\Omega^{*} \neq
0$, remarkably a unique fixed point is found in both forms of the
model, and moreover it takes on a reasonable simple form. The specific
results are:
\begin{itemize}
\item \emph{Fixed points of the original Selkov model} 
\begin{eqnarray}
s_{1}^{*} &=& \frac{ [ \nu_{2} k_{-3} + \nu_{1} 
k_{3} ]}{k_{1} k_{2} k_{3}}\,
\left( k_{2} + k_{-1} \right) \left( 1 - \frac{\nu_1}
{k_2} \right)^{-1}\,, \ \ \ s_{2}^{*} = \frac{\nu_1}
{\nu_2}\,, \nonumber \\ \nonumber \\
a^{*} &=& \frac{\nu_{1} k_{3}}{ [ \nu_{2} 
k_{-3} + \nu_{1} k_{3} ]}\,\left( 1 - 
\frac{\nu_1}{k_2} \right)\,, \ \ \ 
b^{*} = \frac{\nu_1}{k_2}\,. 
\label{fp_Selkov}
\end{eqnarray}

\item \emph{Fixed points of the modified Selkov model}

The trivial fixed point: 
$s_{1}^{*} = 1\,, \ \ s_{2}^{*} = 0\,, \ \ a^{*} = 1\,.$

The non-trivial fixed point:
\begin{eqnarray}
s_{1}^{*} &=& \frac{ [ \nu_{2} k_{-3} + \nu_{1} 
k_{3} ] \left( k_{2} + k_{-1} \right)}{\Delta}\,, 
\ \ \ s_{2}^{*} = \frac{\nu_1}{\nu_2}\,\frac{k_{1}
k_{2} k_{3}}{\Delta}\,\left( 1 - \frac{\nu_1}
{k_2} \right)\,, \nonumber \\ \nonumber \\
a^{*} &=& \frac{\nu_{1} k_{3}}{ [ \nu_{2} 
k_{-3} + \nu_{1} k_{3} ]}\,\left( 1 - 
\frac{\nu_1}{k_2} \right)\,, \ \ \ 
b^{*} = \frac{\nu_1}{k_2}\,, 
\label{fp_modified_Selkov} 
\end{eqnarray}
where
\begin{displaymath}
\Delta \equiv k_{1} k_{2} k_{3} \left( 1 - 
\frac{\nu_1}{k_2} \right)\left( 1 + \frac{\nu_1}
{\nu_2} \right) + \left[ \nu_{2} k_{-3} + 
\nu_{1} k_{3} \right] \left( k_{2} + 
k_{-1} \right) \,.
\end{displaymath}
\end{itemize}

To make contact with the discussion in earlier sections, let us
introduce the notation $\phi_{1}=s_{1}, \phi_{2}=s_{2}, \phi_{3}=a,
\phi_{4}=b$. Then the mean-field equations
(\ref{first})-(\ref{fourth}) take the form $d\phi_{i}/d\tau=f_{i}
(\underline{\phi})$ where $i=1,\ldots,4$. Performing a linear
stability analysis gives the matrix $M_{ij}$. The explicit forms for
the entries of this matrix for the case when $\underline{\phi}^{*}$ is
the fixed point (\ref{fp_modified_Selkov}) are given in the Appendix.

To study the fluctuations we introduce new continuous variables
$x_{1}, x_{2}, x_{3}, x_{4}$ in place of the previously used discrete
variables $m_{1}, m_{2}, n_{1}, n_{2}$, to describe the probability
distribution. The explicit form of the replacements are
\begin{eqnarray}
\frac{m_1}{N_1} &=& \phi_{1} + \frac{x_1}{\sqrt{N_1}}\,, \ \ \  
\frac{m_2}{N_1} = \phi_{2} + \frac{x_2}{\sqrt{N_1}}\,, \nonumber \\  
\frac{n_1}{N_2} &=& \phi_{3} + \frac{x_3}{\sqrt{N_2}}\,, \ \ \  
\frac{n_2}{N_2} = \phi_{4} + \frac{x_4}{\sqrt{N_2}}\,.
\label{replacements_2}
\end{eqnarray}
Proceeding as in section \ref{analysis} we carry out a system-size
expansion. We find the deterministic equations
(\ref{first})-(\ref{fourth}) to leading order and the Fokker-Planck
equation (\ref{FP_eqn}) at next-to-leading order, with $A_{i}
(\underline{x})$ given by Eq.~(\ref{AandM}), except now that $j$ runs
from 1 to 4. So in this case, the probability distribution at
next-to-leading order, $\Pi(\underline{x}, \tau)$, is completely
determined by two $4 \times 4$ matrices $M$ and $B$. The stability
matrix for the fixed point (\ref{fp_modified_Selkov}) is given in the
Appendix, where we also give the entries of the matrix $B$. The
analysis of the Fokker-Planck equation, the conversion to a set of
Langevin equations and the subsequent Fourier analysis is now exactly
as in Section \ref{analysis}, but with $i,j=1,\ldots,4$.

We have calculated the power spectra for the ATP and ADP concentration
fluctuations exactly. We find that these power spectra have very large
and sharp peaks for a wide range of parameter values (Figures 6c and
6d) -- indicating that the concentrations of these molecules will
undergo violent oscillatory behavior within a small-system
setting. Explicit stochastic simulations of the reaction network show
that indeed large amplitude cycling occurs (Figures 6a and 6b). The
constant dashed lines are the predictions from the corresponding
deterministic theory -- showing a complete absence of cycling, as
expected.  The power spectrum peaks found in this example are rather
large -- of the order of 150 for ADP. This indicates that
intracellular environments composed of up to 25,000 ADP molecules
operating according to this reaction scheme will show pronounced
oscillations, caused by amplification of the underlying stochasticity
of the reaction kinetics.

\section{Discussion and conclusions}
\label{concl}
We have described a mechanism by which a microscopic biochemical
system can, loosely speaking, ``beat the central limit theorem'' via
an amplification of stochastic fluctuations in the reaction kinetics
of its constituents. This mechanism leads to sizable oscillations in
the concentrations of the reagents for reaction schemes, which when
modeled using rate equations, show no cycling behavior for any values
of their rate constants. We have illustrated the effect in two very
different biological examples -- self-regulation of a gene, relevant
to the study of circadian rhythms, and the dynamics of ADP, ATP, and
PFK1 concentrations during glycolysis.

We stress here that the system-size expansion always leads to linear
equations for the fluctuations, with coefficients related to the
steady-state concentrations predicted from the first-order theory
(i.e. the deterministic rate equations).  Thus, the evaluation of the
power spectra is simply an exercise in linear algebra. The signature
of the amplification mechanism is the existence of a peak in the power
spectra. The existence of a peak is guaranteed if the stability matrix
of the deterministic rate equations has complex eigenvalues, with
negative real parts to ensure stability. This provides a simple test
for the existence of the mechanism directly from the deterministic
theory. Crudely speaking, if the approach to the deterministic
steady-state occurs via damped oscillations, then the inclusion of
second-order fluctuations will lead to the amplification of sustained
oscillations. It is also important to point out that the mechanism
described here requires no external tuning of rate constants. This is
because the underlying stochasticity has a flat spectrum in frequency
space (i.e. white noise), and this automatically excites the resonant
frequencies of the system.  The oscillations are also robust -- in the
two examples given in this paper, cycles are present over a very broad
range of parameter values.

The examples of circadian rhythms and glycolysis given here are very
simple and designed to illustrate the amplification mechanism in two
well-known areas of cell biology. It will be important to analyze more
realistic (and hence, more complex) reaction networks using the same
techniques. There is intense current interest in analyzing the
dynamical properties of such networks, especially with regard to
robustness [21]. Deterministic rate equations are used for
these analyzes, but this may need to be revisited on the basis of the
results of this paper. For a network involving $M$ different chemical
reagents, the analysis of fluctuations in the system size expansion
reduces to linear algebra with matrices of rank $M$. If oscillatory
behavior is found, the ``special frequencies'' will be closely related
to the $M$ normal modes of the system. This implies an interesting
analogy with the theory of small oscillations in mechanics
[22], whereby complex oscillatory motion can be decomposed
into normal modes, with the lowest frequency modes almost completely
characterizing the dynamics.  As we have stressed, this mechanism only
operates within a microscopic system, such as a cell, or internal
region of a cell. There is, however, a means by which this mechanism
can lead to macroscopic oscillations in a population of cells; namely,
synchronization of the individual cellular cycles through weak
intercellular interactions.

The amplification mechanism described here should be considered as a
possible cause for oscillatory phenomena in more complicated
biological situations. The hypothesis is theoretically elegant in that
it does not require \emph{ad hoc} nonlinearities postulated simply to
``generate'' cycles.  For example, endogenous circadian rhythms in
\emph{eucaryotes} appear to be controlled by a complex transcriptional
feedback loop involving the genes \textit{clock, cry, Bmal1} and
members of the \textit{per} gene family
[23,24]. Oscillations in the activities of these genes may
be explained by this new mechanism if the number of mRNAs,
translational complexes, and product proteins involved are small
enough to allow the amplification term $R$ to dominate. This is not
unreasonable given that the number of mRNAs (for a given gene) in a
eucaryotic cell ranges from $10^{1}-10^{4}$ [25].  This
amplification mechanism may also explain oscillations within the more
complex reality of glycolysis.  Intracellular F6P concentration in the
yeast \textit{Saccharomyces cerevisiae} is approximately 0.1 mM
[26].  Therefore, a cell of 10 $\mu$m diameter contains on
the order of $10^6$ to $10^7$ F6P molecules. In mammalian cells, the
concentration of PFK1 appears to vary between 0.1 and 1 micromolar
[27], yielding on the order of $10^3$ to $10^4$ PFK tetramers
per cell.  Most eucaryotic cells have on the order of $10^9$ ATP
molecules, but some orders of magnitude less ATP participates directly
in glycolysis. Therefore, the system size of glycolysis, especially if
compartmentalized in a eucaryotic cell, approaches that required for
noise amplification to have a pronounced effect.

The condition for this effect of amplified oscillations is simply the
existence of damped oscillations in the corresponding rate equation
model (\emph{i.e.} complex eigenvalues with negative real part in the
stability matrix). Thus, this mechanism is likely to be widespread in
biochemical networks, implying that dynamics in the cytoplasmic
environment may be far richer than formerly supposed on the basis of
chemical rate equations.

\vspace{2.0cm}

\noindent
We thank Stuart Lindsay for introducing us to the subject of circadian
rhythms in procaryotes.  We acknowledge the NSF and NIH for partial
support, under grants DEB-0328267 and DMS/NIGMS-0342388.

\newpage

\renewcommand{\theequation}{\Alph{section}\arabic{equation}}
\setcounter{section}{0} \setcounter{equation}{0}

\begin{appendix}

\section{Explicit forms of the matrices $M$ and $B$}
\label{appA}

As explained in Section \ref{selkov} of the main text, the matrix $M$,
for the set of deterministic equations 
$d\phi_{i}/d\tau = f_{i} (\underline{\phi})$, is defined by 
$M_{ij} = \partial f_{i}/\partial \phi_{j}|_{\rm FP}$. For the system 
(\ref{first})-(\ref{fourth}) the entries of the matrix are
\begin{eqnarray}
M_{11} &=& - \nu_{1} - k_{1} a - k_{-1} b\,, \ \
M_{12}  =  - \nu_{1} - k_{-1} b\,, \nonumber \\
M_{13} &=&  - \sigma^{1/2} \left[ k_{1} s_{1} \right]\,, \ \  
M_{14}  =  \sigma^{1/2}\left[ \Omega k_{-1}\right]\,, \nonumber \\ 
\nonumber \\
M_{21} &=& - k_{2} b - k_{-3} a\,, \ \ 
M_{22}  =  - (\nu_{2} + k_{3}) + (k_{3} - k_{-3}) a
+ (k_{3} - k_{2}) b\,, \nonumber \\
M_{23} &=& \sigma^{1/2} \left[k_{-3} \Omega + 
k_{3} s_{2}\right]\,, \ \ 
M_{24}  = \sigma^{1/2} \left[ k_{2} \Omega + 
k_{3} s_{2} \right]\,, \nonumber \\ \nonumber \\
M_{31} &=& \sigma^{1/2} \left[ (k_{-3} - k_{1}) a - 
(k_{2}+k_{-1}) b \right]\,, \nonumber \\
M_{32} &=&  \sigma^{1/2} \left[ k_{3} + (k_{-3} - 
k_{3}) a - (k_{-1}+ k_{2}+k_{3}) b \right]\,, 
\nonumber \\
M_{33} &=& \sigma \left[ - k_{1} s_{1} - k_{3} s_{2} - 
k_{-3} \Omega \right]\,, 
\ \ M_{34} = \sigma \left[ (k_{-1}+k_{2}) \Omega - 
k_{3} s_{2} \right] \,, \nonumber \\ \nonumber \\
M_{41} &=& \sigma^{1/2} \left[ k_{1} a + k_{-1} b +
k_{2} b \right]\,, \ \ 
M_{42}  =  \sigma^{1/2} \left[ (k_{-1}+k_{2}) b \right]\,, 
\nonumber \\
M_{43} &=& \sigma \left[ k_{1} s_{1} \right]\,,
\ \ M_{44} = \sigma \left[ - (k_{-1}+k_{2})\Omega \right]\,,
\nonumber \\
\label{matrix_elements_M}
\end{eqnarray}
where $\Omega = 1 - s_{1} - s_{2}$. The $a, b, s_{1}$ and $s_2$ in these 
matrix elements are all assumed to be evaluated at the fixed point 
(\ref{fp_modified_Selkov}).

The noise-correlation matrix, $B_{ij}$, is symmetric and given by
\begin{eqnarray}
B_{11} & = & \nu_{1}(1-s_{1}-s_{2}) 
+ k_{1}s_{1}a + k_{-1}(1-s_{1}-s_{2})b\,, \nonumber \\
B_{12} & = & 0\,, \ \ B_{13} = \sigma^{1/2} \left[ k_{1}s_{1}a + 
k_{-1}(1-s_{1}-s_{2})b \right]\,, \ \ B_{14} = - B_{13}\,, 
\nonumber \\ \nonumber \\
B_{22} & = & \nu_{2}s_{2}+k_{2}(1-s_{1}-s_{2})b
+k_{3}s_{2}(1-a-b) + k_{-3}(1-s_{1}-s_{2})a\,, \nonumber \\
B_{23} & = & \sigma^{1/2} \left[ k_{2}(1-s_{1}-s_{2})b
-k_{3}s_{2}(1-a-b) - k_{-3}(1-s_{1}-s_{2})a \right]\,, 
\nonumber \\
B_{24} & = &  \sigma^{1/2} \left[ - k_{2}(1-s_{1}-s_{2})b \right]\,, 
\nonumber \\ \nonumber \\
B_{33} & = & \sigma \left[ k_{1}s_{1}a + 
k_{-1}(1-s_{1}-s_{2})b + k_{2}(1-s_{1}-s_{2})b \right]. 
\nonumber \\
& & \left. +k_{3}s_{2}(1-a-b)+k_{-3}(1-s_{1}-s_{2})a \right]\,, 
\nonumber \\
B_{34} & = & \sigma \left[ -k_{1}s_{1}a - 
k_{-1}(1-s_{1}-s_{2})b - k_{2}(1-s_{1}-s_{2})b \right]\,, 
\nonumber \\ \nonumber \\
B_{44} & = & \sigma \left[ k_{1}s_{1}a + k_{-1}
(1-s_{1}-s_{2})b+k_{2}(1-s_{1}-s_{2})b \right]\,. 
\label{matrix_elements_B}
\end{eqnarray}

\end{appendix}

\newpage


\begin{enumerate}
\item B. Hess, A. Boiteux, \emph{Ann. Rev. Biochem.} {\bf 40}, 237 (1971).
\item A. Goldbeter, S. R. Caplan, \emph{Ann Rev. Biophys. Bioeng.} {\bf 5},
449 (1976).
\item A. Goldbeter, \emph{Biochemical Oscillations and Cellular Rhythms}
(Cambridge University Press, 1996).
\item J. J. Tyson, in \emph{Computational Cell Biology}, C. P. Fall,
E. S. Marland, J. M. Wagner, J. J. Tyson, Eds. (Springer-Verlag, 2002)
Ch. 9.
\item P. L. Lakin-Thomas, S. Brody, \emph{Ann. Rev. Microbiol.} 
{\bf 58}, 489 (2004).
\item M. Nakajima, \emph{et al}, \emph{Science} {\bf 308}, 414 (2005).
\item P. Smolen, \emph{J. Theor. Biol.} {\bf 174}, 137 (1995).
\item J. Keener, J. Sneyde, \emph{Mathematical Physiology} (Springer-Verlag,
1998).
\item Z. Hou, H. Xin, \emph{J. Chem. Phys.} {\bf 119}, 11508 (2003).
\item M. Falcke, \emph{Adv. Phys.} {\bf 53}, 255 (2004).
\item H. Li, Z. Hou, H. Xin, \emph{Phys. Rev. E} {\bf 71}, 061916 (2005).
\item U. Kummer, \emph{et al}, \emph{Biophys. J.} {\bf 89}, 1603 (2005).
\item N. G. Van Kampen, \emph{Stochastic Processes in Physics and
Chemistry} (Elsevier, Amsterdam, 1981).
\item A. J. McKane, T. J. Newman, \emph{Phys. Rev. E} {\bf 70}, 041902 (2004).
\item A. J. McKane, T. J. Newman, \emph{Phys. Rev. Lett.} 
{\bf 94}, 218102 (2005).
\item G. Kurosawa, A. Mochizuki, Y. Iwasa, \emph{J. Theor. Biol.} {\bf 216},
193 (2002).
\item D. T. Gillespie, \emph{J. Comp. Phys.} {\bf 22}, 403 (1976).
\item E. Renshaw, \emph{Modelling biological populations in space
and time} (Cambridge University Press, Cambridge, 1991).
\item S. H. Strogatz, \emph{Nonlinear Dynamics and Chaos} (Addison
Wesley, 1994).
\item E. E. Sel'kov, \emph{Eur. J. Biochem.} {\bf 4}, 79 (1968).
\item G. von Dassow, E. Meir, E. M. Munro, G. M. Odell, \emph{Nature}
{\bf 406}, 188 (2000).
\item H. Goldstein, \emph{Classical Mechanics} (Addison Wesley, Reading, MA,
ed. 2, 1980).
\item D. Bell-{P}edersen, \emph{et al}, \emph{Nature Rev. Genetics} 
{\bf 6}, 544 (2005).
\item M. Akashi, T. Takumi, \emph{Nature Struct. Biol.} 
{\bf 12}, 441 (2005).
\item B. Alberts, \emph{et al}, \emph{Molecular Biology of the Cell} 
(Garland Publishing, New York, ed. 4, 2002).
\item B. Teusink, {\it et al}, \emph{Eur. J. Biochem.} {\bf 267},
1 (2000).
\item J. B. Hansen, C. M. Veneziale, \emph{J. Lab. Clin. Med.} 
{\bf 95}, 133 (1980).
\end{enumerate}

\newpage

\noindent
\textbf{Figure Captions}


\noindent
\underline{Figure 1:} Reaction networks for the two models considered
in this paper. In the gene regulation model (panels a,b), mRNA (M)
and the corresponding enzyme (P) have rates of degradation $\mu_{1}$
and $\mu_{2}$ respectively. Transcription occurs with rate $k_{1}$
(with inhibition), and translation occurs with rate $k_{2}$. The
functional form of inhibition is chosen to be an exponential function
of relative enzyme concentration, with ``regulation parameter''
$\lambda $. In Sel'kov's model for phosphorylation of
fructose-6-phosphate (panels c,d), ATP enters the system with rate
$\nu _{1}$ and ADP leaves the system with rate $\nu _{2}$. The enzyme
PFK1 is activated by ADP, and this complex catalyzes the reaction with
F6P (not shown) and ATP. Crucially, we assume that only one molecule
of ADP is required to activate PFK1.

\vspace{0.3cm}

\noindent
\underline{Figure 2:} A flow diagram indicating the connections
between deterministic and stochastic models of the same reaction
scheme. Within the deterministic model, if limit cycles are absent,
then the model predicts that cycling is absent.  The stochastic model
is analyzed using the system size expansion. At leading order the
deterministic theory is retrieved. The corrections account for
stochastic fluctuations. If these are amplified (with a factor $R$),
then the model predicts large cycles, so long as the typical number of
molecules $N$ satisfies $N^{1/2} \sim R$.

\vspace{0.3cm}

\noindent
\underline{Figure 3:} In panel (a), the gene regulation reaction
network (Figures 1a and 1b) is recast in terms of two baths. Available
units of space (or resources) for mRNA in bath 1 are represented by
constituents $E_{1}$. Likewise, available units for the enzyme in bath
2 are represented by constituents $E_{2}$. Selkov's model for
phosphorylation of F6P (Figures 1c and 1d) is similarly recast in
terms of two baths in panel (b).  ATP ($S_{1}$) and ADP ($S_{2}$)
molecules, along with available units $E_{1}$ are contained in bath 1,
while the PFK1 enzyme ($E_{2}$), and its complexes ($A$ and $B$) are
contained in bath 2.

\vspace{0.3cm}

\noindent
\underline{Figure 4:} Dynamics of mRNA and enzyme concentrations in
the simple gene regulation model defined in Figs. 1a and 1b. The rate
constants are $k_{0}=1.0, k_{2}=0.1, \mu_{1}=\mu_{2}=0.001$, the
regulation parameter is $\lambda=100.0$, and the bath sizes are
$N_{1}=N_{2}/2=20480$.  Panels (a) and (b) show time series of the
relative concentrations of mRNA ($m$) and enzyme ($p$)
respectively. The constant dashed lines are the predictions from the
deterministic theory.  Panels (c) and (d) show the power spectra
(normalized so that $P(0)=1$) associated with the fluctuations in the
mRNA and enzyme concentrations respectively. The resonance peaks are
the signature of this cycling mechanism. The relative amplification of
fluctuations is $R=27.9$ for mRNA, and $R=3.4$ for the enzyme.

\vspace{0.3cm}

\noindent
\underline{Figure 5:} The unnormalized power spectrum for mRNA
fluctuations, as a function of frequency (for parameter values, refer
to Figure 4c).  The solid line is the exact result
(\ref{power_defn_2}), while the circles correspond to the power
spectrum computed from the discrete Fourier transforms of 10,000 mRNA
time series generated from Gillespie's algorithm.

\vspace{0.3cm}

\noindent
\underline{Figure 6:} Dynamics of ATP and ADP concentrations in the
stochastic reformulation of Sel'kov's reaction scheme. The rate
constants are $k_{1}=1.0, k_{-1}=0.5, k_{2}=0.2, k_{3}=0.2,
k_{-3}=1.0, \nu_{1}=0.0005, \nu_{2}=0.03$, and the bath sizes are
$N_{1}=N_{2}=4096$.  Panels (a) and (b) show time series of the
relative concentrations of ATP ($s_{1}$) and ADP ($s_{2}$)
respectively. The constant dashed lines are the predictions from the
deterministic theory.  Panels (c) and (d) show the normalized power
spectra associated with the fluctuations in the ATP and ADP
concentrations respectively. The sharp resonance peaks are the
signature of this cycling mechanism. The relative amplification of
fluctuations is $R=15.1$ for ATP, and $R=150.4$ for ADP.

\newpage

\begin{figure}
\includegraphics[width=5.0in]{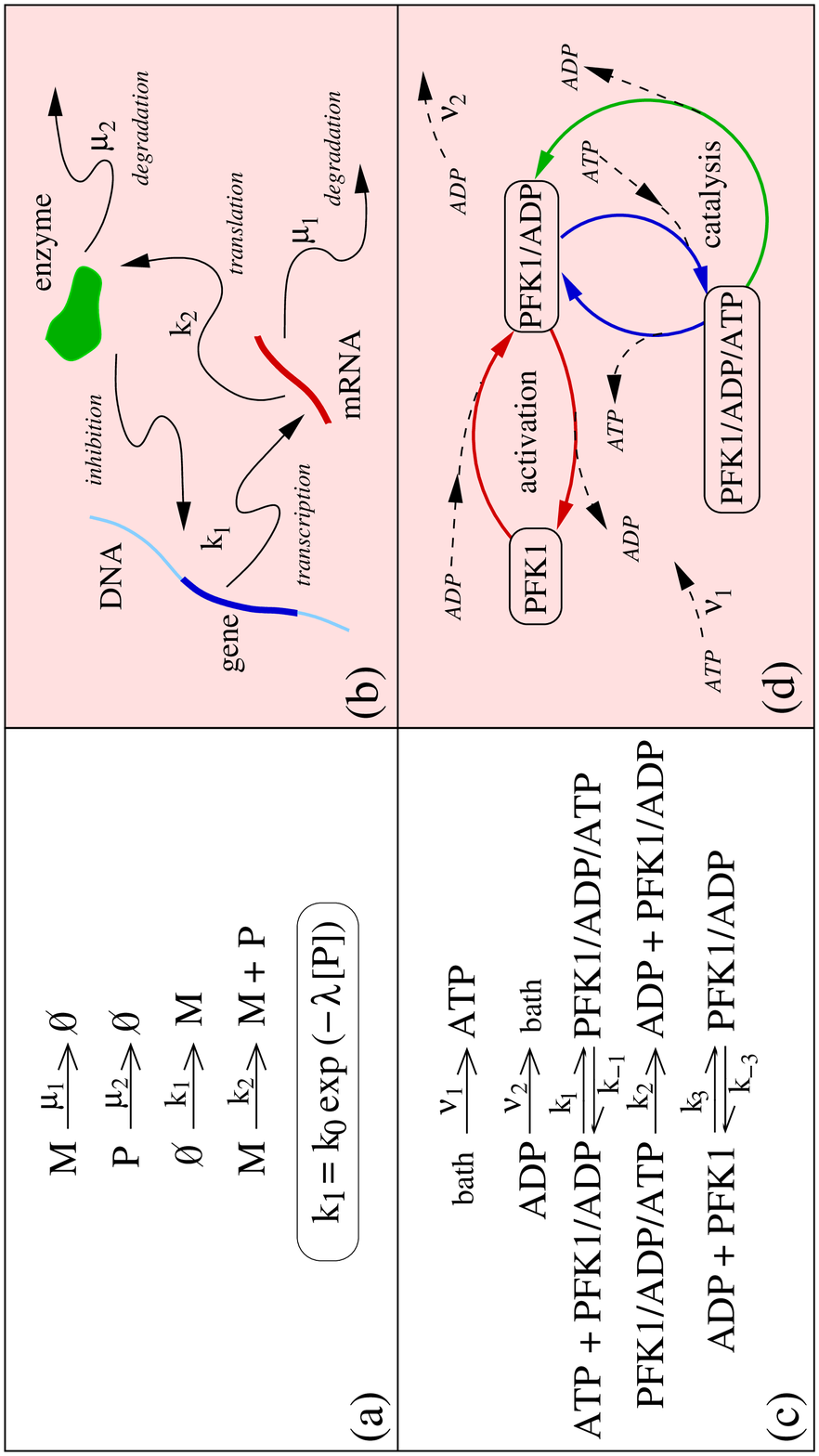}
\caption{}
\end{figure}

\begin{figure}
\includegraphics[width=5.0in]{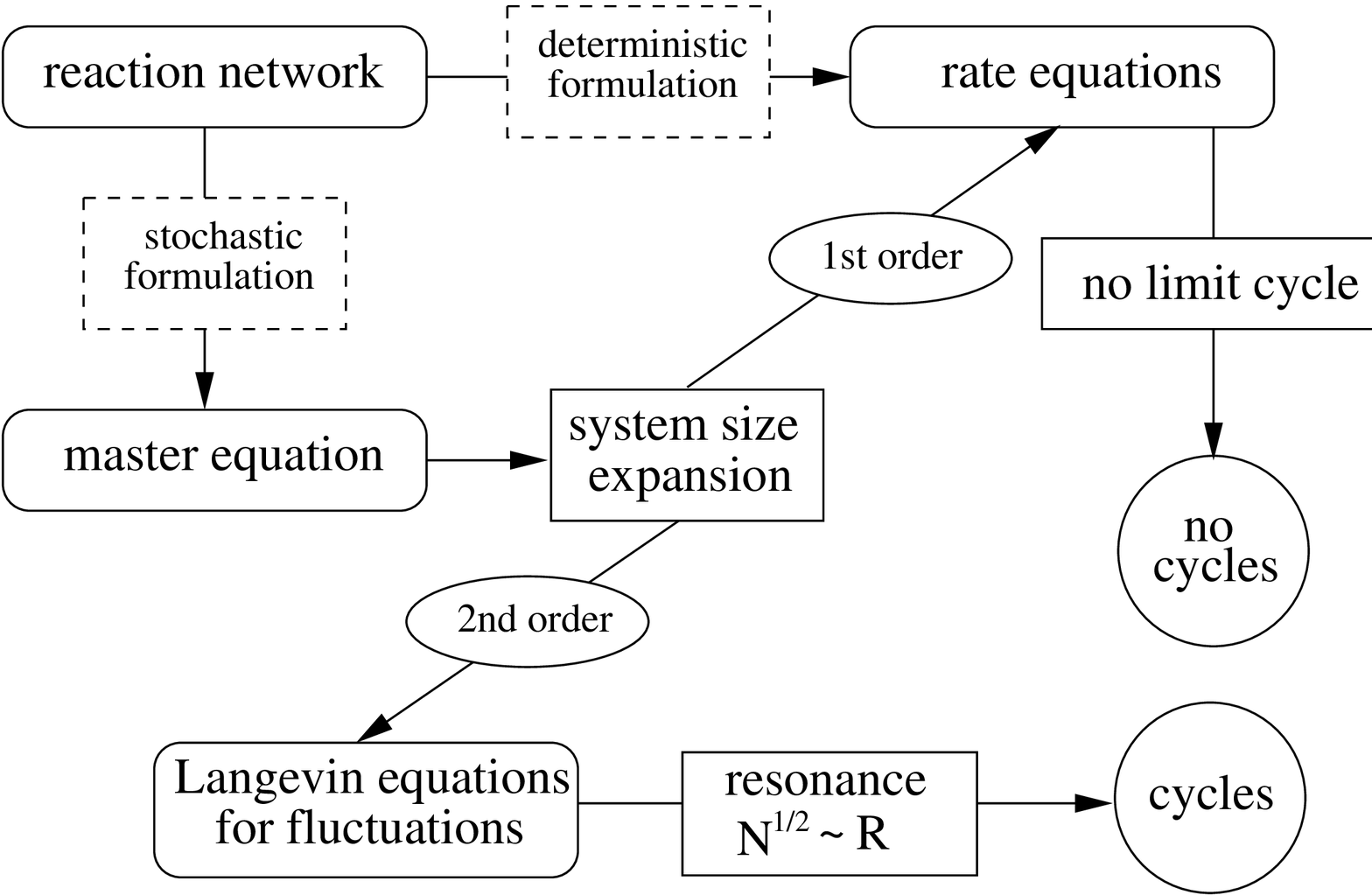}
\caption{}
\end{figure}

\begin{figure}
\includegraphics[width=5.5in]{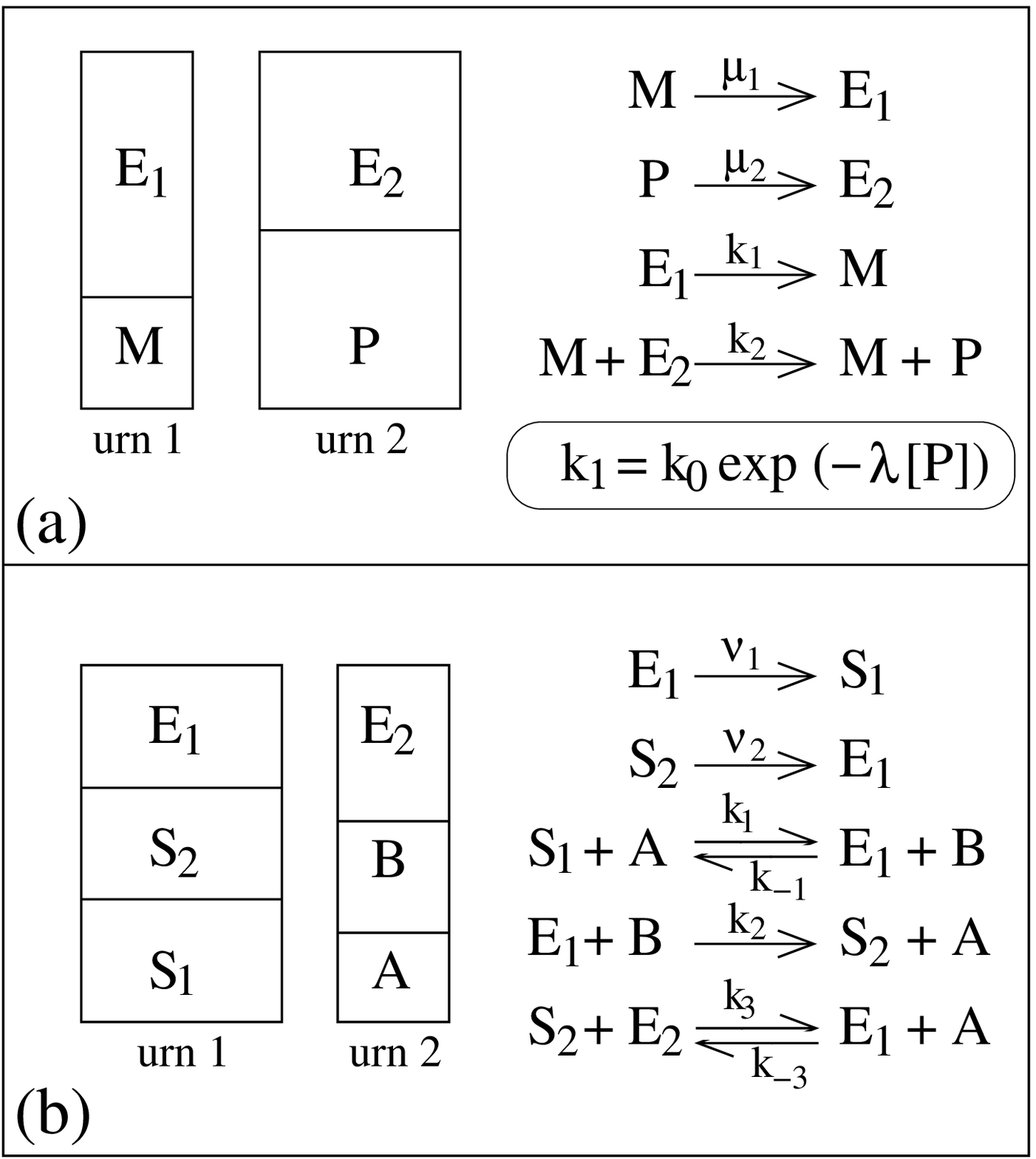}
\caption{}
\end{figure}

\begin{figure}
\includegraphics[width=5.5in]{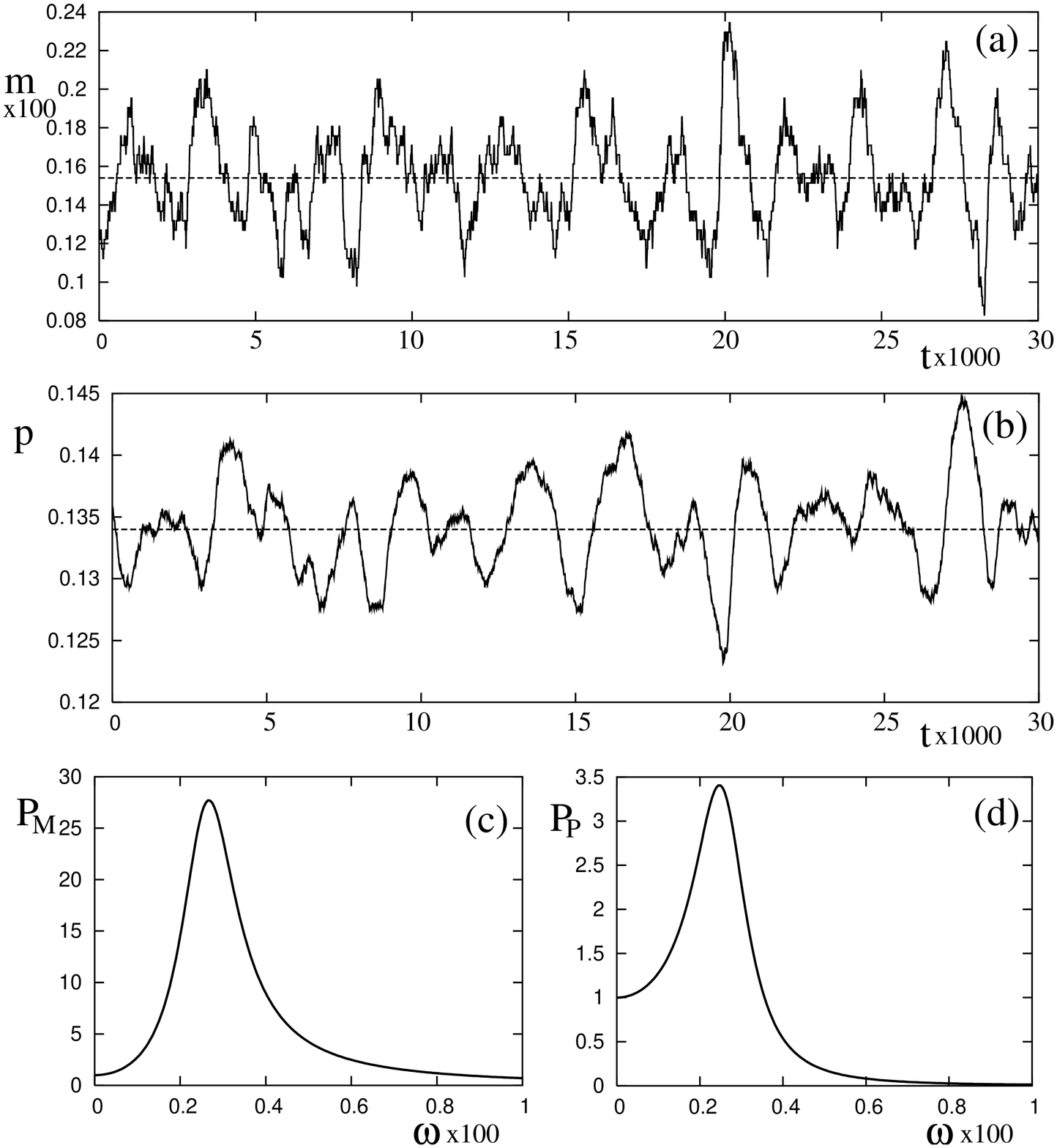}
\caption{}
\end{figure}

\begin{figure}
\begin{center}
\includegraphics[width=5.5in]{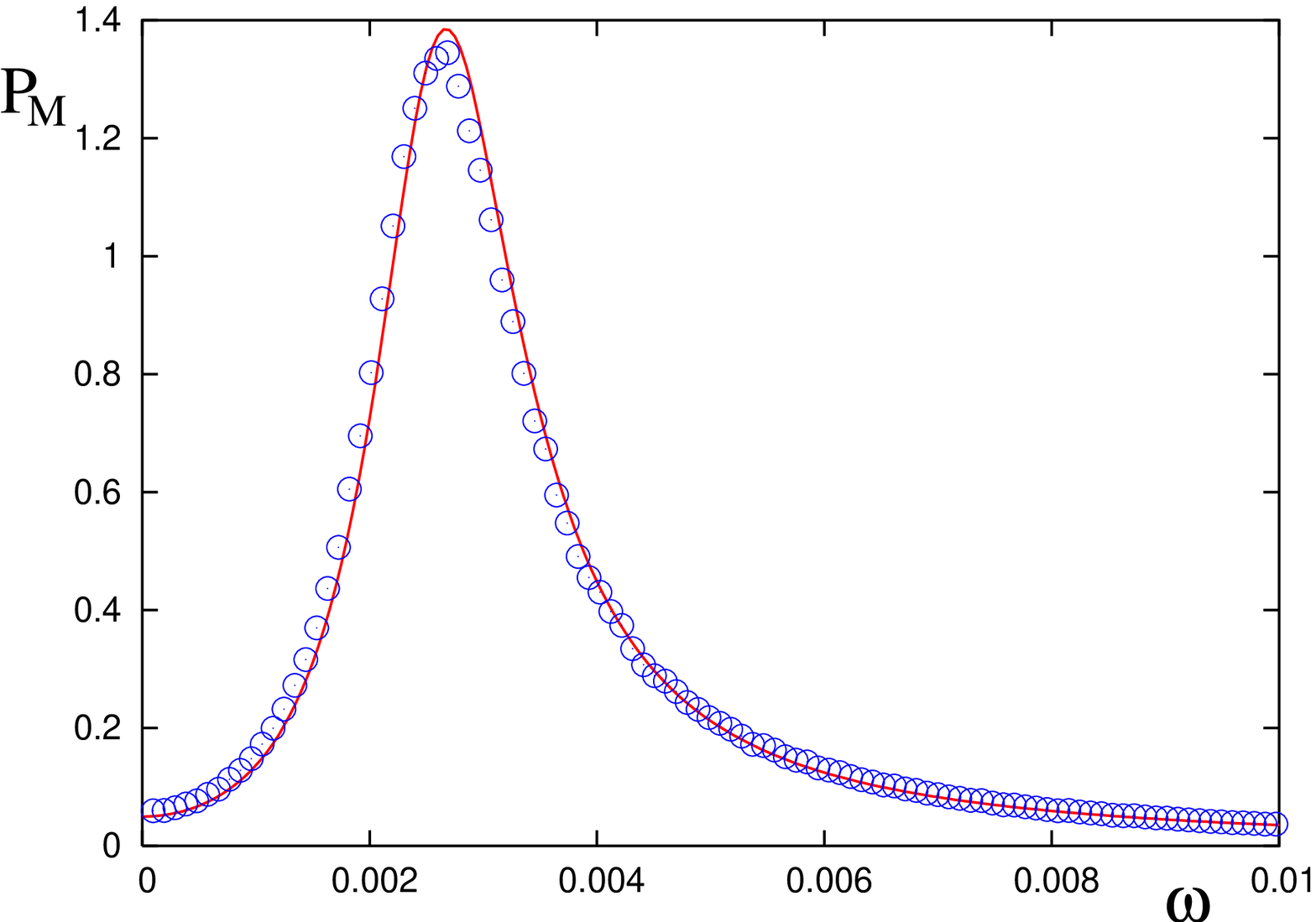}
\caption{}
\end{center}
\end{figure}

\begin{figure}
\begin{center}
\includegraphics[width=5.5in]{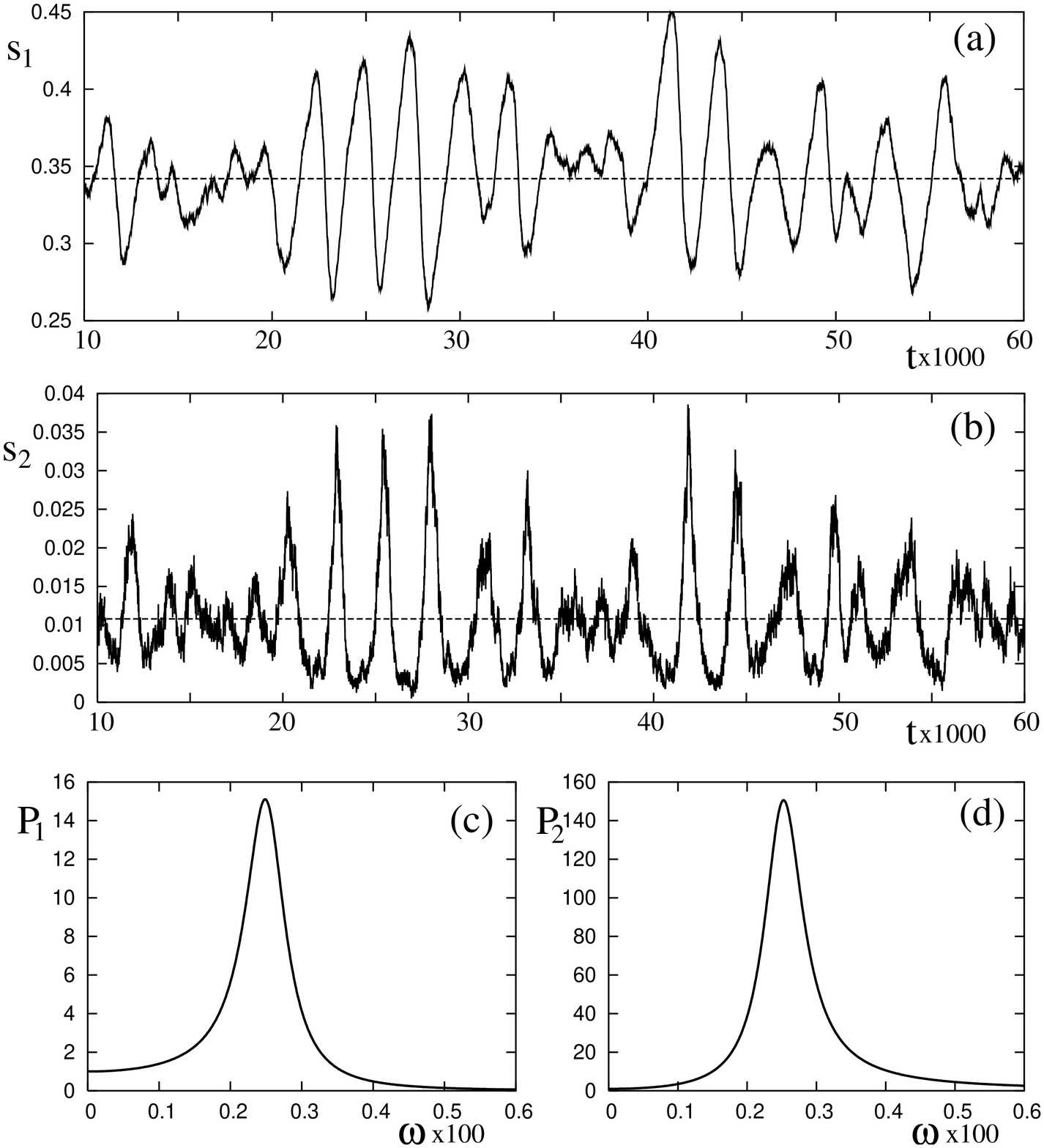}
\caption{}
\end{center}
\end{figure}

\end{document}